\documentclass[journal, twoside]{IEEEtran}
\usepackage{amsmath,amsfonts}
\usepackage{algorithmic}
\usepackage{algorithm}
\usepackage{array}
\usepackage[caption=false,font=footnotesize]{subfig}
\usepackage{textcomp}
\usepackage{stfloats}
\usepackage{graphicx}
\usepackage{url}
\usepackage{verbatim}
\usepackage{graphicx}
\usepackage{cite}
\usepackage{amsmath}
\usepackage{multirow,fixltx2e}
\usepackage{booktabs}
\usepackage{amssymb}
\usepackage{multicol}
\usepackage{tikz}
\usetikzlibrary{positioning, arrows.meta}
\usetikzlibrary{calc}
\usepackage{pgfplots}
\usetikzlibrary{patterns}
\usepackage[flushleft]{threeparttable}

\begin{document}

\title{Video Coding with Cross-Component Sample Offset}

\author{Han~Gao,
  Xin~Zhao,~\IEEEmembership{Senior Member,~IEEE},
  Tianqi~Liu,
  and~Shan~Liu,~\IEEEmembership{Fellow,~IEEE}
  \thanks{Manuscript received XX XX, 2024; revised XX XX, 2024. (\textit{Corresponding author: Han Gao})}
  \thanks{Han~Gao, Xin Zhao, Tainqi Liu, and Shan~Liu are with Tencent America, Palo Alto, California, USA (e-mail: \{gaohan, xinzzhao, tainqitliu, shanl\}@global.tencent.com)).}}

\markboth{IEEE TRANSACTIONS ON IMAGE PROCESSING,~Vol.~XX, No.~XX, XX~2024}%
{Gao \MakeLowercase{\textit{\emph{et al.}}}: Video Coding with Cross-Component Sample Offset}


\maketitle

\begin{abstract}
  Beyond the exploration of traditional spatial, temporal and subjective visual signal redundancy in image and video compression, recent research has focused on leveraging cross-color component redundancy to enhance coding efficiency. Cross-component coding approaches are motivated by the statistical correlations among different color components, such as those in the Y'CbCr color space, where luma (Y) color component typically exhibits finer details than chroma (Cb/Cr) color components. Inspired by previous cross-component coding algorithms, this paper introduces a novel in-loop filtering approach named Cross-Component Sample Offset (CCSO). CCSO utilizes co-located and neighboring luma samples to generate correction signals for both luma and chroma reconstructed samples. It is a multiplication-free, non-linear mapping process implemented using a look-up-table. The input to the mapping is a group of reconstructed luma samples, and the output is an offset value applied on the center luma or co-located chroma sample. Experimental results demonstrate that the proposed CCSO can be applied to both image and video coding, resulting in improved coding efficiency and visual quality. The method has been adopted into an experimental next-generation video codec beyond AV1 developed by the Alliance for Open Media (AOMedia), achieving significant objective coding gains up to 3.5\,\% and 1.8\,\% for PSNR and VMAF quality metrics, respectively, under random access configuration. Additionally, CCSO notably improves the subjective visual quality.
\end{abstract}

\begin{IEEEkeywords}
  Video coding, in-loop filter, Alliance for Open Media (AOMedia), AV1, next-generation video coding, cross-component sample offset (CCSO)
\end{IEEEkeywords}

\section{Introduction}
\label{sec:intro}
\IEEEPARstart{I}{n} recent years, the development of new-generation video coding standards by several standardization organizations has made substantial progress. In 2018, the AOMedia Video 1 (AV1) video coding standard ~\cite{av1-overview1, av1-overview2, av1-overview3, av1_spec} was released, with substantial bitrate reduction achieved over its predecessor VP9 ~\cite{zhao2020comparative}. The latest Versatile Video Coding~(VVC)~\cite{bross2021overview, vvc-spec} standard has been released by the Joint Video Experts Team (JVET) of ITU-T SG 16 WP 3 and ISO/IEC JTC1/SC29/WG11 in July, 2020. The VVC standard was reported to outperform High Efficient Video Coding (HEVC)~\cite{hevc-overview, hevc-spec} by 25\,\% and 35\,\% Bj{\o}ntegaard Delta rate (BD-rate)~\cite{BDrate-2001} savings under All-Intra (AI) and RA configuration~\cite{zhao2020comparative}, respectively, using the traditional PSNR objective quality metrics. The Audio and Video Coding Standard 3 (AVS3)~\cite{zhang2019recent} was finalized in 2019 by the digital audio and video coding standard workgroup of China. In 2021, AOMedia started research activities on exploring next-generation video codec with significantly higher compression efficiency than AV1. A software experimental test model, known as the AOMedia Video Model (AVM) has been developed and used for evaluating input technical contributions related to this exploration activity.

The current mainstream video coding schemes still follow a hybrid block-based processing framework. The prediction mode and residual coding mode for each coding block are usually determined through Rate-Distortion Optimization (RDO). Quantization of transform block coefficients introduces loss of information, causing visual artifacts such as blockiness and ringing.
The quantization step is controlled by the quantization parameter (QP) value, and it varies from 0 to 255 in AVM. The larger the quantization step, the stronger the artifacts appear in reconstructed images.


To effectively reduce the artifacts and improve the visual quality of the reconstructed image, in-loop filters are commonly used in the modern video coding standards. Deblocking filter (DBF)~\cite{norkin2012hevc, karczewicz2021vvc, chen2018overview} is a well-developed filter to address blocking artifacts in AVC, HEVC, VVC, and AV1. The DBF applies a low-pass filter to smooth out the block boundaries and reduce the discontinuities. Additionally, the quantization of transform coefficients also introduces other artifacts, such as blurriness and ringing. In HEVC, Sample Adaptive Offset (SAO)~\cite{fu2011sample} is used to address these artifacts by classifying reconstructed pixels into different categories and applying an offset to each category. The offset is entropy coded into bitstream to enhance the texture. In the latest VVC standard, an Adaptive Loop Filter (ALF)~\cite{karczewicz2021vvc} method is introduced to further reduce these artifacts after applying DBF and SAO. ALF utilizes Wiener filter to reduce the distortion between the source signals and the reconstructed frames. In AV1, new loop filtering tools are also employed to reduce these artifacts after DBF, including the Constrained Directional Enhancement Filter (CDEF)~\cite{midtskogen2018av1} and Loop Restoration Filter (LRF)~\cite{mukherjee2017switchable}. CDEF is capable of identifying the edge direction in a block and applying a high-degree non-linear filter to enhance the edges. LRF is composed of a symmetric Wiener filter and a self-guided filter with subspace projection. Besides those loop filtering methods introduced above, many other in-loop filtering tools~\cite{ding2019cnn, ma2020mfrnet, shao2022ptr, liu2024texture} have also been studied in recent years to further enhance the video quality.

In recent years, cross-component video coding techniques have achieved impressive coding efficiency improvements in many modules within the hybrid block-based video coding framework, including prediction~\cite{lee2009intra}, transform~\cite{misra2019cross}, residual coding~\cite{rudat2019inter}, and loop filtering~\cite{zhao2021study}.
The motivation of cross-component approaches may be explored by the color spaces and chroma sampling basics. The well-know Y'CbCr color format is linearly transformed from RGB color format. Hence, a strong correlation between the Y, Cb, and Cr color components can be easily identified. The Y component contains luma information which has finer structures and details than chroma components. Moreover, in the popular 4:2:0 Y'CbCr sampling format, the chroma components are downsampled by a factor of two in both horizontal and vertical dimension. As a result, each chroma component contains one quarter number of samples in the luma component. Therefore, luma component typically contains richer information than chroma components. Previous works on cross-component video coding technology are mainly motivated by the aforementioned observations.

This paper introduces a novel approach called Cross-Component Sample Offset (CCSO), which employs co-located and neighboring luma samples to derive correction signals for both luma and chroma reconstructed samples. CCSO is characterized as a multiplication-free, non-linear mapping process, implemented via a look-up table. The process takes reconstructed luma samples as input and produces offset values applied to central luma or co-located chroma samples. Experimental results indicate that the proposed CCSO method enhances both image and video coding, leading to superior coding efficiency and visual quality.

The remaining of the paper is organized as follows: Section~\ref{sec:cc coding tools} reviews the related studies on cross-component video coding technology and loop filtering from both academia and industry. Section~\ref{sec:proposed method} details the proposed CCSO technology. Section~\ref{sec:implementation} discusses the encoder implementation and software optimization of CCSO. Section~\ref{sec:exeprimental res} shows the experimental results including coding performance summary, ablation studies, and visual quality comparison. Finally, Section~\ref{sec:conclusion} concludes the work.

\section{Cross-Component Coding Techniques} \label{sec:cc coding tools}
\subsection{Cross-Component Prediction}
Cross Component Linear Mode (CCLM)~\cite{lee2009intra} is a well-studied cross-component prediction tool leveraging inter-channel statistical redundancy. This prediction mode utilizes a linear model to predict chroma samples based on the reconstructed luma samples. The predicted chroma samples, denoted as $\text{pred}_{C}(i,j)$, are computed using
\begin{equation}
  \text{pred}_{C}(i,j) = \alpha \cdot \text{rec}_{L} (i,j) + \beta,
\end{equation}
where $\text{rec}_{L}(i,j)$ represents the downsampled co-located reconstructed luma samples, $i$ and $j$ denote the coordinate of the sample to be predicted, $\alpha$ and $\beta$ denote parameters of the linear mode derived using neighboring chroma samples and the corresponding downsampled luma samples. These parameters are implicitly derived at both the encoder and decoder. In~\cite{zhang2013chroma}, a CCLM mode with three linear models is proposed, and the linear model derivation depends whether left, above or both left and above neighboring samples are used to derive the linear model. In a subsequent work~\cite{zhang2016improving}, an adaptive template selection method was proposed, enabling chroma prediction not only by the Y component but also by the Cb component or an adaptive combination of Y and Cb. Furthermore, an improved CCLM approach is proposed in~\cite{li2022adaptive}, which is based on univariate polynomial regression to model available adjacent samples of chroma blocks and predict chroma samples using reconstructed samples in co-located luma blocks. In~\cite{zhang2017md}, a Multi-Directional Linear Model was proposed including left and top versions. These advancements enable more efficient chroma intra prediction by leveraging both cross-component correlation and spatial correlation among image samples.

In~\cite{zhang2017multi}, a multi-model CCLM (MM-CCLM) approach is introduced, which incorporates multiple linear models within a coding block. In MM-CCLM, reconstructed neighboring luma and chroma samples of the current block are classified into several groups, and different linear models may be derived for different groups. Additionally, in~\cite{li2022jvetz051} and~\cite{zhang2018enhanced}, an LM-angular prediction (LAP) method is proposed, which combines traditional intra-prediction and MM-CCLM prediction using a weighted sum as the prediction samples, exploiting cross-component and spatial correlations simultaneously to enhance prediction efficiency. Furthermore, in~\cite{zhang2018enhanced}, a multi-filter CCLM method is introduced, where four additional candidate down-sampling filters are employed on top of the CCLM prediction mode, and the best filter is selected and signaled to downsample luma samples for deriving the linear model parameters and performing cross-component prediction.

In~\cite{yeo2011chroma}, template matching was utilized for chroma prediction by employing the reconstructed luma block. Meanwhile, a chroma from luma (CfL) prediction mode~\cite{trudeau2018predicting} is employed in AV1, and the prediction block is derived as the sum of the chroma DC contribution and the scaled luma AC contribution. The DC contribution of a block is the average value of the block, while the AC contribution is derived by removing the DC contribution from the block. In CfL mode, the model parameters, such as the scaling factor applied to the luma AC contribution, are calculated during the encoding process and signaled into the bitstream. Notably, signaling scaling parameters in the bitstream enables encoder-only fitting of the linear model, reducing decoder complexity and enhancing prediction accuracy. In \cite{Astola2022jvetz064}, a convolutional cross-component model (CCCM) is proposed, which predicts a chroma sample using multiple downsampled luma samples located within a cross-shape filter centered at the co-located luma sample. The filter coefficients used in CCCM mode are computed via least mean square minimization between the reconstructed chroma and predicted chroma samples in the reference area of the predicted unit.

\subsection{Cross-Component Residual Coding}
Numerous studies have focused on eliminating cross-component redundancy using linear models for residual pixel values. Kim~\emph{et al.}~\cite{kim2015cross} introduced the Cross Component Prediction (CCP) method, predicting the chroma residual signal from the luma residual signal. CCP dynamically switches predictors based on a linear model to code the residuals of the second and/or the third color components using the residuals of the first color component. This approach subtracts the residuals of the remaining color component ($r_{CR}$) from the residuals of the main color component ($r_{CM}'$) before transform and quantization, employing a linear operation expressed as
\begin{equation}
  r_{CR}' = r_{CR} - \gamma \cdot r_{CM},
\end{equation}
where $\gamma$ is a weighting factor. An extension of this work~\cite{nguyen2015extended} allows the Cb residual to predict the Cr residual. Khairat~\emph{et al.}~\cite{khairat2014adaptive} leveraged the correlation between residual components in 4:4:4 format with CCP, predicting the second and the third components from the first component in RGB and Y'CbCr color spaces, with the model parameters signaled in the bitstream.

In~\cite{bross2021developments}, a Joint Coding of Chroma Residual (JCCR) coding mode was introduced and utilized in VVC standard to further reduce redundancy that exist in the residual signals between two chroma components. Rather than signaling the residuals separately, one of three JCCR modes, featuring various weighting combinations of a single-coded chroma residual, can be selectively applied at the coding unit level. A related method was also proposed in~\cite{rudat2019inter} by applying a block-wise, rotational Inter-Component Transform (ICT) on top of two residual signals from intra or inter prediction of Cb and Cr components. By applying inter-component transformation to joint coding of chroma residual signals, better energy compaction is achieved.

\subsection{Cross-Component Signaling}
For chroma component intra prediction, modern video coding standard, e.g., HEVC and VVC, offers the direct mode (DM)~\cite{hevc-DM}, which utilizes the same intra prediction mode applied to the co-located luma component for the chroma components. This mode effectively reduces signaling overhead for the chroma component, based on the assumption that the texture direction in chroma coding block is similar to that of the co-located luma coding block, allowing reuse of luma intra modes by associated chroma coding blocks. In addition, an enhanced cross-component context modeling technique on top of the AV1 intra mode coding scheme is proposed in~\cite{jin2021improved}, which signals the chroma nominal mode is based on the co-located luma nominal mode. Additionally, leveraging the high correlation between chroma and luma delta angles, the context for chroma delta angles is derived from the delta angles of co-located luma blocks.

\subsection{Cross-Component Loop Filtering}
Misra~\emph{et al.}~\cite{misra2019cross} present a cross component adaptive loop filter (CC-ALF), which leverages the correlation between luma and chroma samples to enhance the quality of chroma samples. It applies a linearly filtered version of co-located luma samples to generate corrections for chroma samples, effectively reducing the compression distortion. This approach extends previous ALF methods by refining chroma samples using filtered luma samples after a single-channel process. Unlike traditional ALF, which handles luma and chroma channels separately, CC-ALF processes them jointly to exploit cross-component relationships. Additionally, the luma filtering occurs concurrently with the single-channel ALF luma filtering, avoiding increased latency and line buffer requirements. The filter coefficients for CC-ALF are determined using a Wiener-filter approach at the encoder side and signaled in the Adaptation Parameter Set (APS). This filtering process, selectively enabled and disabled across the image, aims to refine each chroma component using the corresponding luma component, enhancing overall image quality.

In scenarios where chroma components contain abundant textures and edges, using the reconstructed chroma samples to restore blurred textures in the luma component can improve coding efficiency. Cross-component SAO (CCSAO) has been proposed in~\cite{kuo2022cross}, which utilizes collocated luma samples to classify chroma samples and enhance the quality of reconstructed chroma samples. CCSAO reduces sample distortion by leveraging the strong correlation between luma and chroma components to classify reconstructed samples into different categories and deriving one offset for each category, which is then added to the samples in that category. The offsets for each category are derived at the encoder and signaled in the bitstream. To maintain low complexity, only the band information of reconstructed samples is considered for sample classification in CCSAO.

\section{Proposed Cross-Component Sample Offset}
\label{sec:proposed method}
\subsection{Motivation}

\begin{figure}[t]
  \centering
  \centerline{\includegraphics[width=0.49\textwidth]{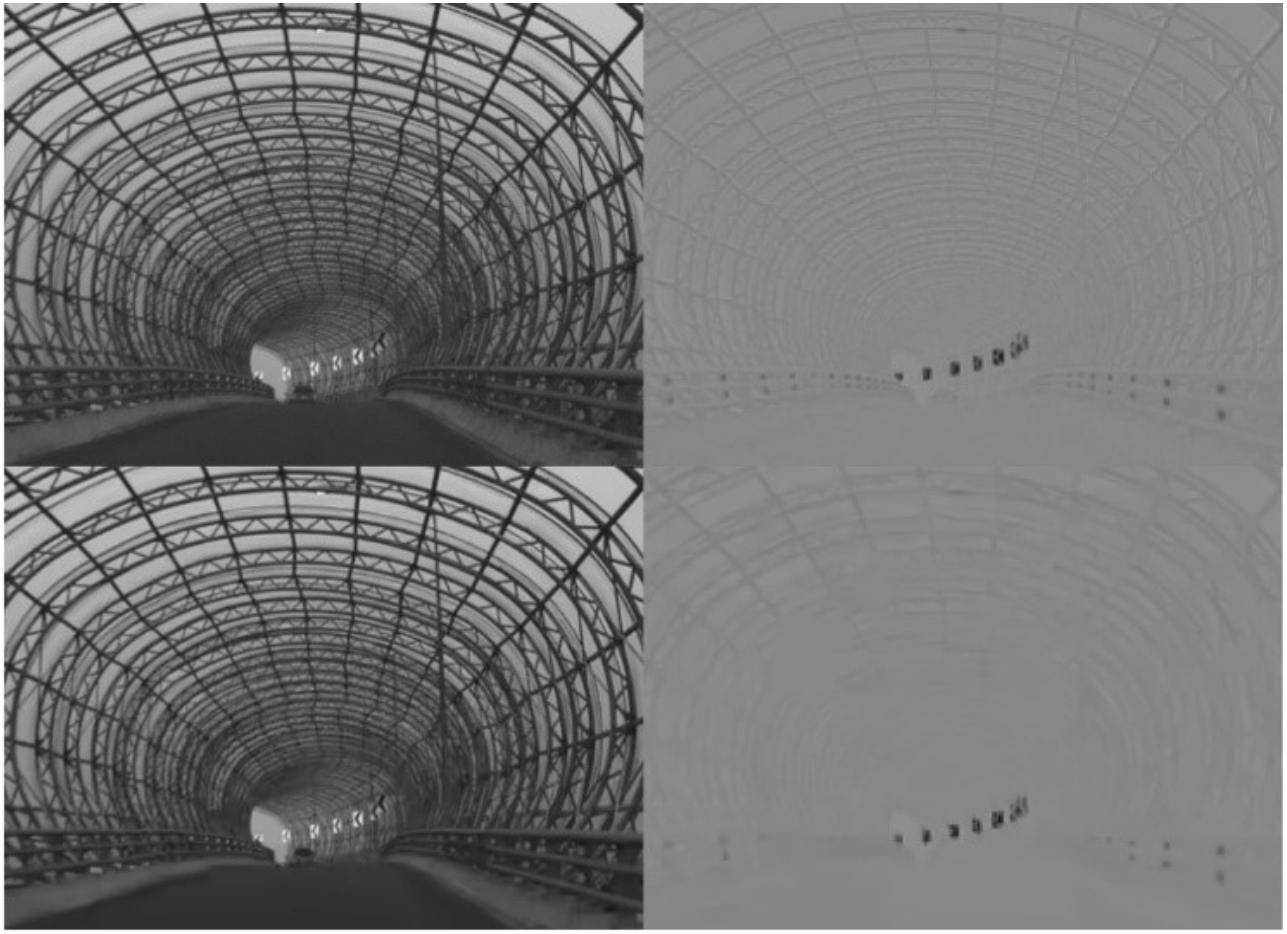}}
  \caption{Comparison of detailed textures between luma and chroma component; top row: original Y (left) and Cb (right) channels, bottom row: reconstructed Y (left) and Cb (right) channels after compression.}
  \label{luma_chroma}
\end{figure}

As illustrated in Fig.~\ref{luma_chroma}, for lossy coding of a camera captured image, the luma channel preserves a significant amount of texture details, whereas the chroma channels suffer greater information loss in structural texture patterns. To exploit the correlations between the luma and chroma channels during the in-loop filtering process, CC-ALF~\cite{misra2019cross} was introduced in VVC. In CC-ALF, filters are applied to the reconstructed luma sample values to enhance the reconstructed chroma sample values, using a typical linear weighted sum of co-located and neighboring luma reconstruction samples. However, the linear relationship assumption of CC-ALF between luma and chroma samples may not always hold true.

To make the cross-component loop filtering more flexible without linear filtering limitation, a CCSO in-loop filtering method is introduced, which is designed to improve both the luma and chroma reconstructed signals using information from the luma reconstructed signal. Different from CC-ALF, CCSO is featured as a non-linear filtering process. In the subsequent subsections, technical design of CCSO will be introduced with details.

\subsection{Filtering Process}
\label{sec:filtering process}
\begin{figure}[t]
  \centering
  \resizebox{0.49\textwidth}{!}{%
    \begin{tikzpicture}[node distance=1cm and 2cm, every node/.style={rectangle, draw, minimum height=1cm, minimum width=0cm, align=center, font=\bfseries\Large, line width=0.7mm}]
    \node (input) [draw=none] {};
    \node (deblocking) [right=of input] {Deblocking};
    \node (cdef) [above right=1cm and 2.5cm of deblocking] {CDEF};
    \node (ccso) [below right=1cm and 2.5cm of deblocking] {CCSO};
    \node (loop) [right=7.5cm of deblocking] {Loop Restoration};
    \node (output) [right=of loop, draw=none] {};

    \draw[->, >=stealth, line width=0.5mm] (input) -- (deblocking);
    \draw[->, >=stealth, line width=0.5mm] (deblocking.east) -- ++(0.75,0) |- (cdef.west);
    \draw[->, >=stealth, line width=0.5mm] (deblocking.east) -- ++(0.75,0) |- (ccso.west);
    \draw[->, >=stealth, line width=0.5mm] (cdef.east) -| ++(1.5,0) |- (loop.west);
    \draw[->, >=stealth, line width=0.5mm] (ccso.east) -| ++(1.5,0) |- (loop.west);
    \draw[->, >=stealth, line width=0.5mm] (loop) -- (output);
\end{tikzpicture}
  }
  \caption{Location of the proposed CCSO filter in the loop filtering pipeline on top of the AV1 codec.}
  \label{fig:ccso_cdef}
\end{figure}
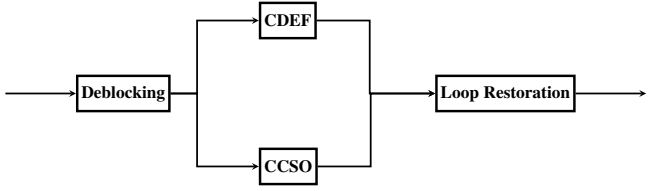

As illustrated in Fig.~\ref{fig:ccso_cdef}, CCSO operates concurrently with CDEF. The reconstructed samples following deblocking are used as input for both CCSO and CDEF. CCSO produces offset values, which are added to the reconstructed samples of the luma and chroma components to reduce reconstruction error. This in-loop filter arrangement offers a better balance between hardware implementation cost and coding efficiency compared to other alternatives.

CCSO is designed for improved loop filtering on both luma and chroma components, and the filtering process of CCSO involves three main steps. First, the current reconstructed luma samples, i.e., the output of the deblocking process, are classified using two types of classifiers: the edge-offset (EO) classifier and the band-offset (BO) classifier. These classifiers can operate jointly or individually based on indicators signaled at the frame level. Second, the class associated with the current luma sample is used as an index to fetch offset values from a lookup table (LUT), which is determined at the frame level with entries selected from a limited number of predefined values. This LUT is shared across the entire frame. Finally, the derived offset values using the LUT and class index are added to the corresponding luma and chroma components. A filter unit-level on/off flag (non-overlapped $256\times256$ luma samples) is signaled to indicate whether CCSO filtering is applied for the associated filter unit.



\subsection{CCSO classifier}

\subsubsection{BO classifier}
The BO classifier utilizes only a single input luma sample. When CCSO is applied to the luma component, the current luma sample is used in the BO classifier. When CCSO is applied to the chroma components, the co-located luma sample serves as the input for the BO classifier.

Additionally, the BO classifier can be used in conjunction with the EO classifier or independently. A frame-level flag is signaled to indicate whether the current frame uses only the BO classifier or combines both the BO and EO classifiers. In both scenarios, the BO index ($\text{BO}_{\text{idx}}$) is derived using the following formula:
\begin{equation}
  \text{BO}_{\text{idx}} = I_{\text{rl}} (x,y) \gg (\text{bitDepth} - {\log_2 N_{\text{band}}}),
  \label{eq:bo_idx}
\end{equation}
where $I_\text{rl}$ denotes the intensity values of the reconstructed luma samples located at $(x,y)$, $\text{bitDepth}$ corresponds to the bit depth of the input signal, and $N_\text{band}$ is the maximum allowed number of bands which are indicated in the frame header. The supported values of $N_\text{band}$ are limited to 1, 2, 4, and 8 when BO and EO are jointly applied. When BO is applied alone, the supported values of $N_\text{band}$ include 1, 2, 4, 8, 16, 32, 64, and 128. The maximum allowed number of bands is signaled at the frame level using two or three bits. The computed $\text{BO}_{\text{idx}}$ from (\ref{eq:bo_idx}) is used in the LUT to derive the offset values of CCSO.

\subsubsection{EO classifier}

\begin{figure}[t]
  \centering
  \centerline{\includegraphics[width=0.23\textwidth]{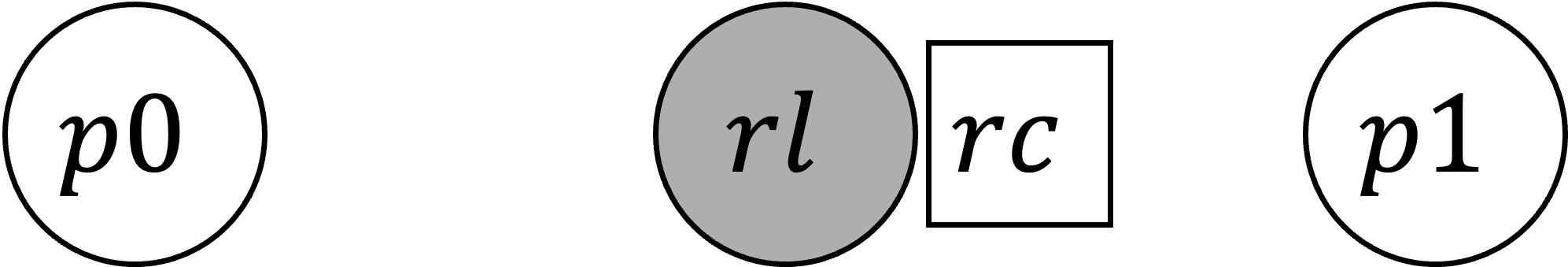}}
  \caption{The input of CCSO EO classifier are three luma reconstructed samples $p_0$, $p_1$ and $rl$; $rc$ is the current chroma sample which applies CCSO, and $rl$ is the co-located luma sample of $rc$.}
  \label{fig_input}
\end{figure}

The EO classifier employs a three-tap filter, utilizing three reconstructed luma samples as input. As illustrated in Fig.~\ref{fig_input}, the current luma sample (when CCSO is applied to the luma component) or the co-located luma sample (when CCSO is applied to the chroma components), denoted as $rl$, along with its neighboring samples $p_0$ and $p_1$, whether adjacent or non-adjacent, are used in this classifier.

\begin{figure}[t]
  \centering
  \centerline{\includegraphics[width=0.4\textwidth]{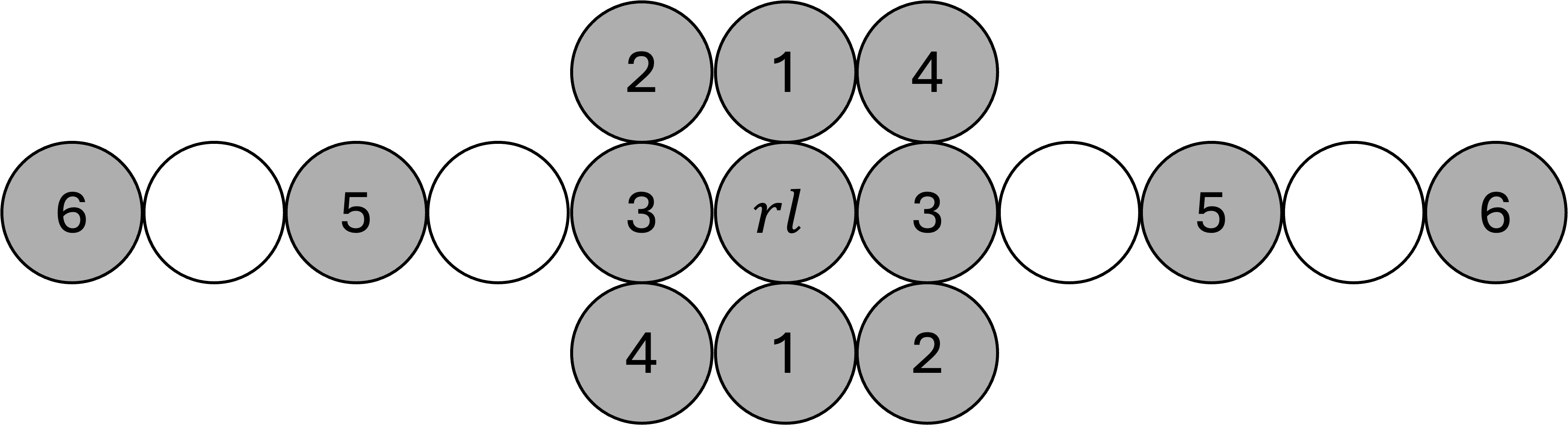}}
  \caption{Six switchable filter shapes of CCSO, one filter shape is selected at frame-level for each color component.}
  \label{fig_filter_shape}
\end{figure}

To better preserve edges in various scenarios, the locations of the neighboring input luma samples are adjustable. As illustrated in Fig.~\ref{fig_filter_shape}, six filter shapes are defined by the positions of $p_0$ and $p_1$, including four with adjacent filter taps (filters 1\textendash4) and two with non-adjacent filter taps (filters 5 and 6). These filter taps are switchable at the frame level, and the selection of the filter shape is indicated by a 3-bit index. This design of filter shapes ensures that only one top and one bottom line of reconstructed samples are involved, thereby minimizing line buffer memory requirements.

\begin{figure}[t!]
  \centering
  \subfloat[EO quantizer \textit{edge\_clf0} with three quantizaiton indices.]{
    \includegraphics[width=0.47\textwidth]{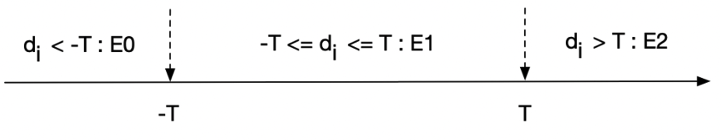} 
  }
  \hfill
  \subfloat[EO quantizer \textit{edge\_clf1} with two quantizaiton indices.]{
    \includegraphics[width=0.47\textwidth]{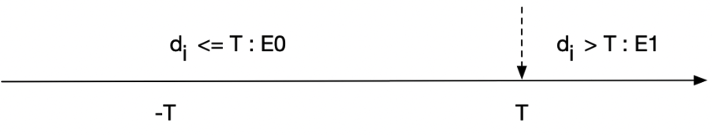} 
  }
  \caption{Illustration of the two proposed EO quantizer types.}
  \label{fig_adaptive_edge}
\end{figure}

Given the intensity values of $p_0$, $p_1$ and $rl$, the EO classifier index $\text{EO}_{\text{idx0,1}}$ as derived as following. The delta values between $p_0$, $p_1$ and $rl$ are computed first, denoted as $m_0$ and $m_1$, respectively, described as follows,
\begin{equation}
  m_0 = p_0 - rl;
\end{equation}
\begin{equation}
  m_1 = p_1 - rl.
\end{equation}
The delta values $m_0$ and $m_1$ are further quantized into two or three indices depending on EO quantizer types. As illustrated in Fig.~\ref{fig_adaptive_edge}, two types of EO quantizers are supported. That is, the \textit{edge\_clf0}, which quantize $m_0$ and $m_1$ into one of three quantization indices using (\ref{eq:edge_clf0}), and \textit{edge\_clf1}, which quantizes them into one of two quantization indices as in (\ref{eq:edge_clf1}),
\begin{equation}
  \text{EO}_{\text{idx0,1}} =
  \begin{cases}
    0, & m_{0,1} < -T,                     \\
    1, & -T \leqslant m_{0,1} \leqslant T, \\
    2, & m_{0,1} > T;
  \end{cases}
  \label{eq:edge_clf0}
\end{equation}
\begin{equation}
  \text{EO}_{\text{idx0,1}} =
  \begin{cases}
    0, & m_{0,1} < -T,         \\
    1, & m_{0,1} \geqslant -T, \\
  \end{cases}
  \label{eq:edge_clf1}
\end{equation}
where $T$ is denoted as quantization step with $T\in\{8, 16, 32, 64\}$.


\subsection{Derivation of Offsets}
A combined CCSO class index $\text{CCSO}_{\text{idx}}$ is calculated using the BO index and the EO indices as follows:
\begin{equation}
  \text{CCSO}_{\text{idx}}= (\text{BO}_\text{idx} \ll 4) + (\text{EO}_\text{idx0} \ll 2) + \text{EO}_\text{idx1}.
  \label{eq:ccso_idx}
\end{equation}
If only BO classifier is used without EO classifier, the $\text{EO}_\text{idx0,1}$ are set as zero. The $\text{CCSO}_{\text{idx}}$ will be used as the input of the offset values LUT to derive the corresponding offset values. This LUT is adaptively generated for individual color component inside one frame at encoder and signaled into the bitstream.

The size of of this LUT depends on the number of possible values of $\text{CCSO}_{\text{idx}}$. For example, when the maximum number of bands is eight, and the EO quantizer type provides three quantization intervals, the LUT size in this case equals to $8 \times 3 \times 3 = 72$, which is the largest possible size for BO and EO classifiers combined case. For BO classifier standalone case, largest possible LUT size is 128, calculated as $128 \times 1 \times 1$.

The entries $s_{i}$ with $i = \text{CCSO}_{\text{idx}}$ of this LUT directly reflect the offset values that will be added on the output of the CDEF. Thus, the values are carefully selected based on experimental results and limited as
\begin{equation}
  s_{i} \in \{0, 1, -1, 3, -3, 7, -7, -10\}.
  \label{eq:offsets}
\end{equation}
At the frame level, a truncated unary syntax element is signaled to indicate each individual $s_{i}$. At the encoder side, the method for determining the best offset value for a certain class is described in Section~\ref{sec:encoder search}.

  \begin{algorithm}[t]
    \label{alg:frame_level_syn}
    \caption{CCSO Syntax Design} 
    \newcommand*\BitNeg{\ensuremath{\mathord{\sim}}}
\begin{algorithmic}
    \STATE \textbf{Syntax}  \hfill \textbf{Descriptor}
    \STATE frame\_header()
    \STATE \textit{\textbf{ccso\_frame\_flag}}   \hfill u(1)${}^\ast$
    \IF {ccso\_frame\_flag} 
    \FOR{plane = 0 to num\_planes}
      \STATE \textit{\textbf{ccso\_enable}}[plane]   \hfill   u(1)
      \IF {ccso\_enable[plane]}
        \STATE \textit{\textbf{ccso\_bo\_only}}[plane]   \hfill   u(1)
        \IF {ccso\_bo\_only[plane]}
          \STATE \textit{\textbf{max\_band\_log2}}[plane]   \hfill   u(3)
        \ELSE
          \STATE \textit{\textbf{max\_band\_log2}}[plane]   \hfill   u(2)
          \STATE \textit{\textbf{quant\_step\_idx}}[plane]   \hfill   u(2)
          \STATE \textit{\textbf{filter\_shape\_idx}}[plane]   \hfill   u(3)
          \STATE \textit{\textbf{edge\_clf}}[plane]   \hfill   u(1)
        \ENDIF
        \STATE max\_band = 1 $\ll$ max\_band\_log2 
        \STATE intervals = ccso\_bo\_only ? 1 : ({\BitNeg}edge\_clf + 2)
        \FOR {d\_0 = 0 to intervals}
          \FOR {d\_1 = 0 to intervals}
            \FOR {bdn = 0 to max\_band}
              \STATE ccso\_idx = bnd $\ll$ 4 + d\_0 $\ll$ 2 + d\_1
              \STATE \textit{\textbf{offset\_idx}}    \hfill tu${}^\dagger$
              \STATE offset\_lut[plane][ccso\_idx] 
              \STATE \quad \quad = offsets[offset\_idx]
            \ENDFOR
          \ENDFOR
        \ENDFOR
      \ENDIF
    \ENDFOR
    \ENDIF
    \end{algorithmic}
    \begin{tablenotes}
      \footnotesize
      \item ${}^\ast$ Fixed-length coded syntax elements.
      \item ${}^\dagger$ Truncated unary coded syntax elements.
    \end{tablenotes}
  \end{algorithm}

\subsection{Applying Offsets}

The offset values are separately derived for luma and chroma component. If CCSO is enabled for luma, the CDEF reconstructed luma samples are corrected by
\begin{equation}
  rl'(x,y) = \text{clip}(rl(x,y) + s_\text{luma} (x,y)),
\end{equation}
where $rl(x,y)$ is the luma reconstructed sample from CDEF, $s_\text{luma} (x,y)$ is the derived CCSO offset values, and $rl'(x,y)$ is the final filtered reconstructed samples by CCSO;
and when CCSO is enabled for Cb and Cr components, the corrected chroma samples are given by
\begin{equation}
  rc'_{\text{cb}}(x,y) = \text{clip}(rc_{\text{cb}}(x,y) + s_\text{cb} (x,y)),
\end{equation}
and
\begin{equation}
  rl'_{\text{cr}}(x,y) = \text{clip}(rc_{\text{cr}}(x,y) + s_\text{cr} (x,y)),
\end{equation}
where $rc_{\text{cb, cr}}(x,y)$ are reconstructed samples of CDEF, $rc_\text{cb,cr} (x,y)$ are CCSO offset values, and $rc'_{\text{cb, cr}}(x,y)$ are final filtered samples in Cb and Cr components, respectively.

Additionally, a fixed-size non-overlapped filter unit with $256\times256$ luma samples, i.e., $128\times128$ chroma samples for 4:2:0 color format, are introduced. The CCSO filter can be turned on or off at this filter unit level for each color component, separately.





\subsection{Syntax Elements Design}
\label{syntax}
The syntax elements signaling of CCSO can be categorized into frame-level and filter-unit-level.


The frame-level signaled syntax elements are given by Algorithm~1, more specifically,
\begin{itemize}
  \item One-bit flag \textit{ccso\_frame\_flag} flag indicates whether CCSO is applied on any color component of the current frame;
  \item One-bit flag \textit{ccso\_enable} indicates whether CCSO is applied in the current component of the current frame;
  \item One-bit flag \textit{band\_only\_flag} indicates if BO classifier is used alone or combined with EO classifier;
  \item Two (or three)-bits syntax \textit{max\_band\_log2} specifies the logarithm base two value of the maximum number of allowed bands.
  \item Two-bits index \textit{quant\_step\_idx} indicates the selection of quantization step size;
  \item Three-bits index \textit{filter\_shape\_idx} indicates the selection of filter shape used in EO classifier;
  \item One-bit flag \textit{edge\_clf} indicates the EO quantizer type;
  \item For each CCSO class, offset values indices \textit{offset\_idx} indicate the LUT entries from eight pre-defined values descried in (\ref{eq:offsets}), which are signaled using truncated unary code (up to 7 bits for each index).
\end{itemize}

At filter-unit-level, a context coded flag is signaled to indicate whether the CCSO filter is enabled or not for a specific filer unit. Three contexts are added for signaling this flag, one for each color component.

\section{Implementation of Proposed Methods}
\label{sec:implementation}
As described in Section~\ref{sec:proposed method}, the proposed CCSO method is a highly adaptive loop filtering technique with multiple syntaxes designed to accommodate various video content. This approach offers a substantial search space for CCSO filtering parameters. To achieve optimal coding gain, it is essential to develop an efficient encoder implementation of the proposed method, alongside an effective Single Instruction/Multiple Data (SIMD) implementation for the filtering process.
\subsection{Encoder Search of CCSO Parameters}
\label{sec:encoder search}
There are seven frame-level and one filter-unit-level CCSO filtering parameters to be determined by encoder, as described in Section~\ref{syntax}. The derivation of a group of five frame-level syntax, including \textit{ccso\_bo\_only},  \textit{filter\_shape\_idx}, \textit{quant\_step\_idx}, \textit{max\_band\_log2}, and \textit{edge\_clf}, involves evaluating different combinations of model parameters through cascaded loops. This approach results in a substantial search space if implemented using a brute-force search method.

For each combination of the five frame-level filtering parameters, the encoder must determine the filter-unit-level CCSO filter enabling flag for each filtering unit and derive the offset LUT. Given that the derivation of the offset LUT is intertwined with determining the filter units applying CCSO, an iterative search strategy is implemented to optimize both the filter-unit-level enabling flags and the LUT entries.

Initially, the filter-unit-level CCSO filter enabling flag is set to one for all filtering units, assuming that CCSO filtering will be applied to all units in the current frame. The optimization process then proceeds iteratively through the following steps:

Step 1: Based on the current determination of filtering units with CCSO enabled, derive the optimal LUT by calculating the difference between the original samples and the reconstruction samples for each combination of CCSO classes ($\text{CCSO}_\text{idx}$) as given by (\ref{eq:ccso_idx}).

Step 2: Using the updated LUT, compare the rate-distortion (R-D) cost of enabling versus disabling CCSO filtering for each filtering unit. Set the CCSO filter enabling flag according to the option with the lower R-D cost.

The above steps 1 and 2 are repeated until a maximum of 15 iterations is reached or the accumulated filter-unit-level R-D cost for the current picture ceases to decrease. According to the determined filter-unit-level CCSO enabling flag, the frame-level CCSO enabling flags \textit{ccso\_enable} for individual component and \textit{ccso\_frame\_flag} for the whole frame are set.


\subsection{SIMD Optimization}
The proposed CCSO loop filtering method is highly compatible with SIMD architecture, and an implementation of SIMD for the CCSO filtering process has been successfully developed and validated to significantly reduce the runtime complexity of the software implementation.

The CCSO loop filtering process consists of two primary steps:

Step 1 involves determining the index for the Lookup Table (LUT) entry, which is calculated based on the BO and EO classes, denoted as the combination of $\text{BO}_\text{idx}$, and $\text{BO}_\text{idx0,1}$.

Step 2 entails retrieving the offset from the LUT and adding this offset value to the reconstruction samples.

In the CCSO encoder design, $\text{BO}_\text{idx0}$ and $\text{BO}_\text{idx1}$ for each reconstruction sample are computed on a frame-level basis before the application of filtering. The SIMD implementation of step one is straightforward, as determining the LUT entry involves simply packing the BO and EO classes indices into a single value. For step 2, the operation of fetching the LUT offset can be efficiently executed using the \textit{\_mm256\_shuffle\_epi8} instruction, taking advantage of the small size of the LUT, which comprises up to 9 entries for individual $\text{BO}_\text{idx0}$, each limited to 3 bits. The runtime saving achieved by the proposed SIMD optimization is shown in Table~\ref{table_ccso_simd}. For the overall encoder runtime, average 1.2\,\% and 1.6\,\% savings are achieved for RA and LD configurations, respectively. For the overall decoder runtime, average 5.0\,\% and 6.2\,\% savings are achieved for RA and LD configurations, respectively.

\begin{table}[t]
  \caption{SIMD Optimization Runtime Comparison (Anchor: w/o SIMD, Test w/ SIMD)}
  \begin{center}
    \scalebox{0.85}{
      \begin{tabular}{lcccccc}
    \toprule
                 & \multicolumn{2}{c}{AI}      & \multicolumn{2}{c}{RA}  & \multicolumn{2}{c}{LD} \\
    \midrule
                 & \textbf{Enc-T}        & \textbf{Dec-T}          & \textbf{Enc-T}          & \textbf{Dec-T}     & \textbf{Enc-T}        & \textbf{Dec-T}      \\
    \midrule
    A1$\_$4k      & 101.6\,\%          & 96.4\,\%          & 99.9\,\%          & 93.8\,\%   & $-$ & $-$  \\
    A2$\_$2k      & 99.5\,\%          & 93.8\,\%          & 98.8\,\%          & 94.7\,\%       & 98.5\,\%          & 93.2\,\%      \\
    A3$\_$720p    & 98.6\,\%          & 92.8\,\%          & 98.4\,\%          & 94.6\,\%    & 98.0\,\%          & 94.8\,\%         \\
    A4$\_$360p    & 97.9\,\%          & 96.2\,\%          & 98.3\,\%          & 98.2\,\%        & 97.8\,\%          & 92.6\,\%   \\
    A5$\_$270p    & 98.9\,\%          & 101.2\,\%          & 98.6\,\%          & 95.3\,\%    & 98.1\,\%          & 93.4\,\%            \\
    B1$\_$SYN     & 99.9\,\%          & 94.7\,\%          & 98.5\,\%          & 95.0\,\%   & 99.1\,\%          & 95.0\,\%        \\
    \textbf{AVG}  & \textbf{99.5\,\%} & \textbf{94.9\,\%} & \textbf{98.8\,\%} & \textbf{95.0\,\%} & \textbf{98.4\,\%} & \textbf{93.8\,\%} \\
    \bottomrule
  \end{tabular}
    }
    \label{table_ccso_simd}
  \end{center}
\end{table}

\section{Experimental Results}
\label{sec:exeprimental res}
\subsection{Test Condition}
The proposed algorithm has been implemented based on tag v6.0.0 of AVM reference software\cite{AVM_research_tag}. The common test condition (CTC)~\cite{AV2_CTC} specified by AOM Testing Subgroup is used to conduct the experiment. The test set has a total of 56 sequences, including camera captured sequences with resolutions from 4K to 240p (class A1{\textendash}A5), synthetic content sequences (class B1), and screen content sequences (class B2). The QP settings are from 110 to 235 for RA and LD, and from 85 to 210 for AI. BD-Rate is used to evaluate the coding gain, which is calculated for each color component (Y/Cb/Cr) using PSNR and VMAF. The overall BD-rate is also reported using 14/16, 1/16, and 1/16 weightings on the quality scores of Y, Cb, and Cr components (marked as YCbCr in the results tables), respectively. The complexity is measured by the ratio between the encoding and decoding time of the test and anchor.





\subsection{Coding Performance and Complexity Analysis}

\begin{table*}[t]
  \caption{Overall Coding Performance of CCSO under the AI, RA, and LD Configurations}
  \begin{center}
    \scalebox{1.2}{
      \begin{tabular}{clcccccccc}
  \toprule
                             &                     & \textbf{Y}         & \textbf{Cb}        & \textbf{Cr}        & \textbf{YCbCr}     & \textbf{VMAF(Y)}   & \textbf{Enc-Time} & \textbf{Dec-Time} \\
  \midrule
  \multirow{8}*{\textbf{AI}} & A1\_4k              & $-$0.08\%          & $-$2.24\%          & $-$3.88\%          & $-$0.48\%          & $-$0.42\%          & 107\%             & 109\%             \\

                             & A2\_2k              & $-$0.16\%          & $-$4.20\%          & $-$4.68\%          & $-$0.62\%          & $-$0.71\%          & 104\%             & 107\%             \\

                             & A3\_720p            & $-$0.29\%          & $-$4.01\%          & $-$4.17\%          & $-$0.65\%          & $-$0.58\%          & 102\%             & 105\%             \\

                             & A4\_360p            & $-$0.25\%          & $-$9.31\%          & $-$5.87\%          & $-$0.88\%          & $-$1.26\%          & 101\%             & 109\%             \\

                             & A5\_270p            & 0.02\%             & $-$2.23\%          & $-$1.23\%          & $-$0.10\%          & $-$0.50\%          & 102\%             & 107\%             \\

                             & B1\_SYN             & $-$0.45\%          & $-$4.58\%          & $-$4.93\%          & $-$0.88\%          & $-$1.10\%          & 104\%             & 109\%             \\

                             & \textbf{AVG w/o B2} & \textbf{$-$0.22\%} & \textbf{$-$4.37\%} & \textbf{$-$4.41\%} & \textbf{$-$0.64\%} & \textbf{$-$0.75\%} & \textbf{104\%}    & \textbf{107\%}    \\

                             & B2\_SCC             & $-$0.25\%          & $-$1.33\%          & $-$0.93\%          & $-$0.32\%          & $-$1.15\%          & 103\%             & 104\%             \\
  \midrule

  \multirow{8}*{\textbf{RA}} & A1\_4k              & $-$0.40\%          & $-$3.77\%          & $-$5.39\%          & $-$0.91\%          & $-$0.85\%          & 106\%             & 107\%             \\

                             & A2\_2k              & $-$0.32\%          & $-$6.26\%          & $-$5.97\%          & $-$0.85\%          & $-$0.70\%          & 103\%             & 104\%             \\

                             & A3\_720p            & $-$0.27\%          & $-$5.28\%          & $-$5.43\%          & $-$0.75\%          & $-$0.33\%          & 103\%             & 103\%             \\

                             & A4\_360p            & $-$0.19\%          & $-$7.53\%          & $-$6.53\%          & $-$0.79\%          & $-$0.53\%          & 101\%             & 102\%             \\

                             & A5\_270p            & 0.01\%             & $-$1.67\%          & $-$4.04\%          & $-$0.25\%          & $-$1.07\%          & 102\%             & 100\%             \\

                             & B1\_SYN             & $-$0.42\%          & $-$5.43\%          & $-$6.09\%          & $-$0.94\%          & $-$0.78\%          & 103\%             & 104\%             \\

                             & \textbf{AVG w/o B2} & \textbf{$-$0.30\%} & \textbf{$-$5.41\%} & \textbf{$-$5.75\%} & \textbf{$-$0.81\%} & \textbf{$-$0.69\%} & \textbf{103\%}    & \textbf{104\%}    \\

                             & B2\_SCC             & $-$1.87\%          & $-$2.24\%          & $-$2.30\%          & $-$1.92\%          & $-$5.25\%          & 110\%             & 105\%             \\
  \midrule

  \multirow{8}*{\textbf{LD}} & A1\_4k              & $-$                  & $-$                  & $-$                  & $-$                  & $-$                  & $-$                 & $-$                 \\

                             & A2\_2k              & $-$0.62\%          & $-$7.42\%          & $-$7.29\%          & $-$1.24\%          & $-$1.07\%          & 104\%             & 106\%             \\

                             & A3\_720p            & $-$0.33\%          & $-$5.40\%          & $-$4.79\%          & $-$0.78\%          & $-$0.77\%          & 104\%             & 107\%             \\

                             & A4\_360p            & $-$0.02\%          & $-$7.06\%          & $-$5.45\%          & $-$0.55\%          & 0.20\%             & 102\%             & 106\%             \\

                             & A5\_270p            & $-$0.15\%          & $-$4.80\%          & $-$6.21\%          & $-$0.58\%          & $-$1.25\%          & 101\%             & 106\%             \\

                             & B1\_SYN             & $-$0.77\%          & $-$6.33\%          & $-$8.13\%          & $-$1.40\%          & $-$0.84\%          & 103\%             & 106\%             \\

                             & \textbf{AVG w/o B2} & \textbf{$-$0.48\%} & \textbf{$-$6.57\%} & \textbf{$-$6.72\%} & \textbf{$-$1.05\%} & \textbf{$-$0.82\%} & \textbf{103\%}    & \textbf{106\%}    \\
                             & B2\_SCC             & $-$3.22\%          & $-$3.59\%          & $-$4.14\%          & $-$3.28\%          & $-$7.99\%          & 109\%             & 105\%             \\
  \bottomrule
\end{tabular}
    }
    \label{table_ccso}
  \end{center}
\end{table*}

Table~\ref{table_ccso} presents the BD-rate improvement achieved by CCSO across AI, RA, and LD configurations, respectively, wherein the anchor is AVM tag v6.0.0 with CCSO disabled (achieved by setting runtime flag --enable-ccso=0), and the test is AVM tag v6.0.0 with CCSO enabled (achieved by setting runtime flag --enable-ccso=1). A negative number in Table~\ref{table_ccso} indicates coding gain (or bitrate saving under the same quality) when CCSO is applied.

Overall, CCSO demonstrates a promising coding gain, exhibiting improvements of 0.64\,\%, 0.81\,\%, and 1.05\,\% YCbCr coding gain under AI, RA, and LD configurations,respectively. Notably, substantial enhancements are observed in the chroma channels, with notable gains chroma component, e.g., 5.41\,\% and 5.75\,\% Cb and Cr coding gain under RA configuration. It is worth mentioning that for individual sequences, such as \textit{WestWindEasy} in natural video content, the maximum RA coding gains of 0.88\,\%, 26.36\,\%, and 28.83\,\% for the Y, Cb, and Cr components, respectively, are achieved. For screen content sequence \textit{Slides2r} in class B2, the maximum RA coding gains of 3.45\,\%, 13.75\,\%, and 12.16\,\% for Y, Cb, and Cr components, respectively, are observed.

\begin{figure*}[t]
  \centering
  \subfloat[]{%
    \begin{tikzpicture}
    \def \w {6.7cm}
    \def \h {4.1cm}
    \pgfplotsset{tick label style={font=\scriptsize},
    label style={font=\scriptsize},
    legend style={font=\scriptsize},
    }
    \begin{axis}[
        minor y tick num=1,
		height = \h,
		width = \w,
        legend pos= north east,
        scale only axis,
        xmin=0,
        xmax=9000,
        xlabel = bitrate(kbps),
        ymin = 36,
        ymax= 51,
        y tick label style={
            /pgf/number format/.cd,
                fixed,
                fixed zerofill,
                precision=0,
            /tikz/.cd
        },
        ylabel = Cb-PSNR (dB),
        y label style={at={(0.08,0.5)}},
        grid = both,
        legend pos=south east
	]
	\addplot [
        smooth,
		solid,
        mark=triangle*,
        mark options={scale=0.8}
	] coordinates {
        (70.822154, 39.609061)
        (201.193846, 41.782387)
        (594.808615, 44.031668)
        (1643.996308, 46.129088)
        (4060.600615, 47.690512)
        (8732.030769, 49.266191)
    };
	\addplot [
        smooth,
		solid,
        gray,
        mark=square,
        mark options={scale=0.6}
	] coordinates {
		(70.764923, 37.971344)
        (198.696, 40.706639)
        (591.923077, 43.288333)
        (1640.527385, 45.605143)
        (4055.069538, 47.44676)
        (8726.931692, 49.180813)
    };
    \legend{CCSO on, CCSO off}
    \end{axis}

    
\end{tikzpicture}
    \label{subfig:cb_psnr}
  }
  \hfil
  \subfloat[]{%
    \begin{tikzpicture}
    \def \w {6.7cm}
    \def \h {4.1cm}
    \pgfplotsset{tick label style={font=\scriptsize},
        label style={font=\scriptsize},
        legend style={font=\scriptsize},
    }
    \begin{axis}[
            minor y tick num=1,
            height = \h,
            width = \w,
            legend pos= north east,
            scale only axis,
            xmin=0,
            xmax=9000,
            xlabel = bitrate(kbps),
            ymin = 40,
            ymax= 52,
            y tick label style={
                    /pgf/number format/.cd,
                    fixed,
                    fixed zerofill,
                    precision=0,
                    /tikz/.cd
                },
            ylabel = Cr-PSNR (dB),
            y label style={at={(0.08,0.5)}},
            grid = both,
            legend pos=south east
        ]
        \addplot [
            smooth,
            solid,
            mark=triangle*,
            mark options={scale=0.8}
        ] coordinates {
            (8732.030769, 50.759818)
            (4060.600615, 49.112664)
            (1643.996308, 47.795418)
            (594.808615, 46.223594)
            (201.193846, 44.464401)
            (70.822154, 42.567316)
            };
        \addplot [
            smooth,
            solid,
            gray,
            mark=square,
            mark options={scale=0.6}
        ] coordinates {
                (8726.931692, 50.713769)
                (4055.069538, 48.909569)
                (1640.527385, 47.248307)
                (591.923077, 45.51809)
                (198.696, 43.445912)
                (70.764923, 41.197393)
            };
        \legend{CCSO on, CCSO off}
    \end{axis}
    
    
\end{tikzpicture}
    \label{subfig:cr_psnr}
  }
  \caption{BD-rate curves of sequence \textit{WestWindEasy} from class A3 under RA configuration tested using AVM tag v6.0.0; (a) Cb-PSNR vs bitrate curve; (b) Cr-PSNR vs bitrate curve.}
  \label{fig_BDBR}
\end{figure*}
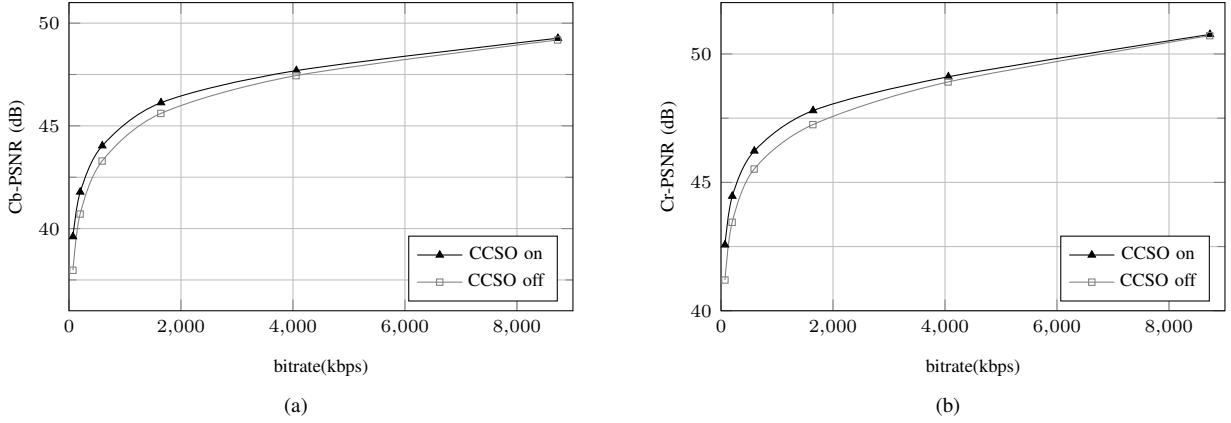

The BD-rate curves for sequence \textit{WestWindEasy} is shown in Fig.~\ref{fig_BDBR}, it is observed that both Cb and Cr channel BD-rate curves show significantly improved coding performance with the proposed CCSO.

When analyzing the performance of CCSO across different video classes, such as class A5 (270p), and class B2 (screen content), it becomes evident that CCSO tends to exhibit relatively different coding gains compared to other test classes. This discrepancy in performance, particularly observed in class A5, may be attributed to the minimal block size required for implementing CCSO. The minimal block size found in class A5 videos contains relatively more intricate video content, which poses challenges for effective offset compensation by CCSO. On the other hand, for the luma channel in the class B2 videos, a relatively high coding gain is observed. This may be due to the screen content videos having less natural noise and clearer structures, which allows CCSO to better exploit the luma offset compensation.

\begin{table}[t]
  \caption{Fixed EO Classifier of CCSO Only on Chroma Components}
  \begin{center}
    \scalebox{1}{
      \begin{tabular}{clccccc}
  \toprule
                             &              & \textbf{Y}        & \textbf{Cb}          & \textbf{Cr}          & \textbf{YCbCr}       \\
  \midrule
  \multirow{7}*{\textbf{AI}} & A1$\_$4k     & 0.10\,\%          & $-$3.51\,\%          & $-$3.64\,\%          & $-$0.33\,\%          \\
                             & A2$\_$2k     & 0.06\,\%          & $-$3.15\,\%          & $-$4.09\,\%          & $-$0.28\,\%          \\
                             & A3$\_$720p   & 0.05 \,\%         & $-$3.45 \,\%         & $-$2.47\,\%          & $-$0.23\,\%          \\
                             & A4$\_$360p   & 0.08\,\%          & $-$6.76\,\%          & $-$9.40\,\%          & $-$0.58\,\%          \\
                             & A5$\_$270p   & 0.10\,\%          & $-$4.37\,\%          & $-$2.55\,\%          & $-$0.21\,\%          \\
                             & B1$\_$SYN    & 0.03\,\%          & $-$2.58\,\%          & $-$2.63\,\%          & $-$0.18\,\%          \\
                             & \textbf{AVG} & \textbf{0.06\,\%} & \textbf{$-$3.57\,\%} & \textbf{$-$3.95\,\%} & \textbf{$-$0.28\,\%} \\
  \midrule

  \multirow{7}*{\textbf{RA}} & A1$\_$4k     & 0.21\,\%          & $-$7.37\,\%          & $-$9.09\,\%          & $-$0.73\,\%          \\
                             & A2$\_$2k     & 0.13\,\%          & $-$6.64\,\%          & $-$8.25\,\%          & $-$0.54\,\%          \\
                             & A3$\_$720p   & 0.09\,\%          & $-$4.67\,\%          & $-$4.44\,\%          & $-$0.33\,\%          \\
                             & A4$\_$360p   & 0.17\,\%          & $-$6.60\,\%          & $-$7.10\,\%          & $-$0.48\,\%          \\
                             & A5$\_$270p   & 0.05\,\%          & $-$6.25\,\%          & $-$1.75\,\%          & $-$0.31\,\%          \\
                             & B1$\_$SYN    & 0.04\,\%          & $-$4.92\,\%          & $-$3.87\,\%          & $-$0.33\,\%          \\
                             & \textbf{AVG} & \textbf{0.12\,\%} & \textbf{$-$6.10\,\%} & \textbf{$-$6.43\,\%} & \textbf{$-$0.47\,\%} \\
  \midrule

  \multirow{7}*{\textbf{LD}} & A1$\_$4k     & $-$                 & $-$                    & $-$           & $-$                                  \\
                             & A2$\_$2k     & 0.04\,\%          & $-$8.84\,\%          & $-$10.84\,\%         & $-$0.84\,\%          \\
                             & A3$\_$720p   & 0.12\,\%          & $-$5.27\,\%          & $-$4.56\,\%          & $-$0.32\,\%          \\
                             & A4$\_$360p   & 0.15\,\%          & $-$7.30\,\%          & $-$7.91\,\%          & $-$0.56\,\%          \\
                             & A5$\_$270p   & 0.16\,\%          & $-$6.89\,\%          & $-$0.62\,\%          & $-$0.17\,\%          \\
                             & B1$\_$SYN    & -0.02\,\%         & $-$7.09\,\%          & $-$5.09\,\%          & $-$0.52\,\%          \\
                             & \textbf{AVG} & \textbf{0.06\,\%} & \textbf{$-$7.56\,\%} & \textbf{$-$7.42\,\%} & \textbf{$-$0.60\,\%} \\
  \bottomrule
\end{tabular}
    }
    \label{table_ccso_edge_chroma}
  \end{center}
\end{table}

Additionally, for B2 classes, the coding gains on chroma channels are relatively lower than natural video contents, except for the class A5. This suggests that the performance of CCSO may vary depending on the content characteristics, with screen content videos demonstrating unique behavior compared to natural video content.

When comparing the relative complexity, excluding class B2, the average encoding time increases by 4\,\% and decoding time by 7\,\% for AI, 3\,\% and 4\,\% for RA, and 3\,\% and 6\,\% for LD configuration, respectively. The effective encoder searching methods and SIMD implementations as described in Section~\ref{sec:encoder search} contributing the relative low complexity of the proposed CCSO method. Upon analysis, there is a more noticeable increase in encoding times under the RA and LD configurations for class B2. This discrepancy underscores the importance of considering the specific video class characteristics when assessing the performance of CCSO.

\begin{table}[t]
  \caption{BO Classifier of CCSO Only on Chroma Components}
  \begin{center}
    \scalebox{1}{
      \begin{tabular}{clccccc}
  \toprule
                             &               & \textbf{Y}        & \textbf{Cb}          & \textbf{Cr}          & \textbf{YCbCr}       \\
  \midrule
  \multirow{7}*{\textbf{AI}} & A1$\_$4k      & 0.09\,\%          & $-$1.45\,\%          & $-$1.62\,\%          & $-$0.12\,\%          \\
                             & A2$\_$2k      & 0.07\,\%          & $-$1.85\,\%          & $-$1.70\,\%          & $-$0.10\,\%          \\
                             & A3$\_$720p    & 0.04\,\%          & $-$1.07\,\%          & $-$0.87\,\%          & $-$0.04\,\%          \\
                             & A4$\_$360p    & 0.06\,\%          & $-$1.53\,\%          & $-$0.47\,\%          & $-$0.03\,\%          \\
                             & A5$\_$270p    & 0.06\,\%          & $-$0.29\,\%          & $-$0.12\,\%          & 0.04\,\%             \\
                             & B1$\_$SYN     & 0.05\,\%          & $-$0.76\,\%          & $-$0.82\,\%          & $-$0.04\,\%          \\
                             & \textbf{AVG}  & \textbf{0.06\,\%} & \textbf{$-$1.34\,\%} & \textbf{$-$1.17\,\%} & \textbf{$-$0.07\,\%} \\
  \midrule

  \multirow{7}*{\textbf{RA}} & A1$\_$4k      & $-$0.17\,\%       & $-$1.82\,\%          & $-$2.01\,\%          & $-$0.37\,\%          \\
                             & A2$\_$2k      & 0.02\,\%          & $-$2.32\,\%          & $-$1.46\,\%          & $-$0.14\,\%          \\
                             & A3$\_$720p    & 0.04\,\%          & $-$2.11\,\%          & $-$0.97\,\%          & $-$0.10\,\%          \\
                             & A4$\_$360p    & 0.03\,\%          & $-$2.41\,\%          & $-$0.90\,\%          & $-$0.11\,\%          \\
                             & A5$\_$270p    & 0.04\,\%          & $-$0.32\,\%          & $-$0.40\,\%          & 0.01\,\%             \\
                             & B1$\_$SYN     & 0.03\,\%          & $-$1.04\,\%          & $-$1.18\,\%          & $-$0.08\,\%          \\
                             & \textbf{AVG}  & \textbf{0.00\,\%} & \textbf{$-$1.84\,\%} & \textbf{$-$1.29\,\%} & \textbf{$-$0.14\,\%} \\
  \midrule

  \multirow{7}*{\textbf{LD}} & A1$\_$4k      & $-$               & $-$                  & $-$                  & $-$                  \\
                             & A2$\_$2k      & 0.06\,\%          & $-$2.56\,\%          & $-$1.65\,\%          & $-$0.12\,\%          \\
                             & A3$\_$720p    & 0.03\,\%          & $-$2.20 \,\%         & $-$1.07\,\%          & $-$0.12\,\%          \\
                             & A4$\_$360p    & 0.10\,\%          & $-$2.71\,\%          & $-$0.92\,\%          & $-$0.05\,\%          \\
                             & A5$\_$270p    & 0.05\,\%          & $-$1.00\,\%          & $-$1.12\,\%          & $-$0.04\,\%          \\
                             & B1$\_$SYN     & 0.03\,\%          & $-$0.98\,\%          & $-$1.87\,\%          & $-$0.10\,\%          \\
                             & \textbf{AVG } & \textbf{0.05\,\%} & \textbf{$-$2.04\,\%} & \textbf{$-$1.49\,\%} & \textbf{$-$0.10\,\%} \\
  \bottomrule
\end{tabular}
    }
    \label{table_ccso_band_chroma}
  \end{center}
\end{table}


\subsection{Ablation Study}
To gain deeper insights into the components of the proposed CCSO method, a comprehensive set of ablation studies has been conducted.

In the first experiment, we isolate the impact of EO classifier (with only EO quantizer type \textit{edge\_clf0}) exclusively on the chroma channel to exploring the use cases of the EO classifier. The results, summarized in Table~\ref{table_ccso_edge_chroma}, reveal notable coding gains across different configurations. Specifically, 0.28\,\%, 0.47\,\%, and 0.60\,\% YCbCr overall BD-rate reduction is observed under AI, RA, and LD configurations. These gains primarily stem from enhancements in the Cb and Cr channels, whereas the Y channel experiences marginal coding because CCSO is only applied to chroma in this test.


Furthermore, we conducted another experiment focusing solely on the BO classifier on chroma components to explain the functionality of BO classifier. The results are detailed in Table~\ref{table_ccso_band_chroma}. 
Overall, BO classifier standalone provides 1.34\,\% and 1.17\,\% coding gain for Cb and Cr components under AI configuration, 1.84\,\% and 1.29\,\% Cb and Cr gains under RA configuration, and 2.04\,\% and 1.49\,\% Cb and Cr gains under LD configuration. The luma coding loss is negligible because CCSO is also not applied on luma component in this test.

Lastly, we conducted an experiment to evaluate the efficiency of applying CCSO exclusively to the luma component. Therefore, a test was designed using both EO and BO classifiers solely on the luma channel. The simulation results, presented in Table~\ref{table_ccso_band_edge_luma}, show significant coding gains for the Y component, with 0.39\,\%, 0.43\,\%, and 0.53\,\% luma BD-rate gain under AI, RA, and LD configurations, respectively.


\subsection{Visual Analysis}
\begin{figure*}[t]
  \centering
  \centerline{\includegraphics[width=\textwidth]{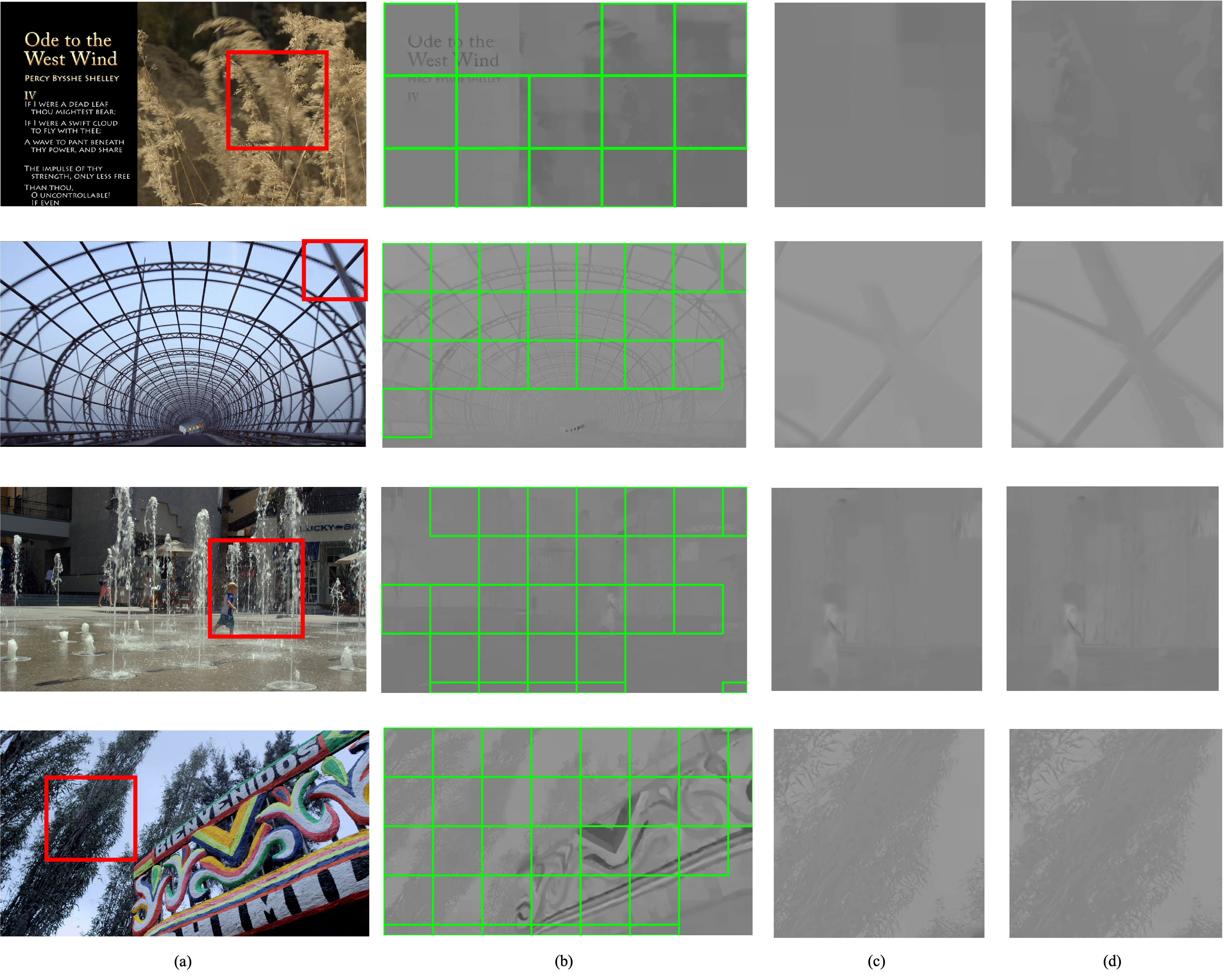}}
  \caption{Visual quality comparison of the proposed CCSO method; (a) original images, 8th frame of \textit{WestWindEasy}, 16th frame of \textit{TunnelFlag}, 1st frame of \textit{ToddlerFountain}, and 1st frame of \textit{Boat}; (b) CCSO applied regions of the corresponding reconstructed Cr component, coded with QP 235 under RA configuration; (c) selected regions (marked with box in original images) with CCSO disabled, (d) selected regions with CCSO enabled.
  }
  \label{visual}
\end{figure*}

\begin{table}[t]
  \caption{EO and BO Classifiers Only on Luma Component}
  \begin{center}
    \scalebox{1}{
      \begin{tabular}{clccccc}
    \toprule
                      &            & \textbf{Y}       & \textbf{Cb}      & \textbf{Cr} & \textbf{YCbCr} \\
    \midrule
    \multirow{7}*{\textbf{AI}} & A1$\_$4k   & $-$0.52\,\% & 0.10\,\% & 0.11\,\%  & $-$0.45\,\%  \\
                      & A2$\_$2k   & $-$0.36\,\% & 0.12\,\% & 0.13\,\%  & $-$0.32\,\%  \\
                      & A3$\_$720p & $-$0.39\,\% & 0.15\,\% & 0.14\,\%  & $-$0.34\,\%  \\
                      & A4$\_$360p & $-$0.43\,\% & 0.17\,\% & 0.18\,\%  & $-$0.38\,\%  \\
                      & A5$\_$270p & $-$0.09\,\% & 0.28,\%  & 0.29\,\%  & $-$0.06\,\%  \\
                      & B1$\_$SYN  & $-$0.42\,\% & 0.09\,\% & 0.09\,\%  & $-$0.37\,\%  \\
                      & \textbf{AVG}& \textbf{$-$0.39\,\%} & \textbf{0.13\,\%} & \textbf{0.13\,\%}  & \textbf{$-$0.34\,\%}  \\
    \midrule

    \multirow{7}*{\textbf{RA}} & A1$\_$4k   & $-$0.76\,\% & 0.07\,\% & 0.21\,\%  & $-$0.67\,\%  \\
                      & A2$\_$2k   & $-$0.41\,\% & 0.07\,\% & 0.08\,\%  & $-$0.37\,\%  \\
                      & A3$\_$720p & $-$0.29\,\% & 0.07\,\% & 0.05\,\%  & $-$0.25\,\%  \\
                      & A4$\_$360p & $-$0.32\,\% & 0.17\,\% & 1.20\,\%  & $-$0.25\,\%  \\
                      & A5$\_$270p & 0.05\,\%  & 0.21\,\%  & 0.50\,\%  & 0.08\,\%  \\
                      & B1$\_$SYN  & $-$0.54\,\% & $-$0.17\,\% & 0.30\,\%  & $-$0.48\,\%  \\
                      & \textbf{AVG} & \textbf{$-$0.43\,\%} & \textbf{0.04\,\%} & \textbf{0.28\,\%}  & \textbf{$-$0.37\,\%}  \\
    \midrule

    \multirow{7}*{\textbf{LD}} & A1$\_$4k   & $-$         & $-$         & $-$        & $-$        \\
                      & A2$\_$2k   & $-$0.61\,\% & 0.13\,\% & 0.37\,\%  & $-$0.54\,\%  \\
                      & A3$\_$720p & $-$0.56\,\% & $-$0.03\,\% & 0.08\,\%  & $-$0.50\,\%  \\
                      & A4$\_$360p & $-$0.36\,\% & 0.05\,\%  & $-$0.41\,\%  & $-$0.35\,\%  \\
                      & A5$\_$270p & $-$0.12\,\% & 0.71\,\% & $-$0.60\,\%  & $-$0.09\,\%  \\
                      & B1$\_$SYN  & $-$0.59\,\% & 0.25\,\% & $-$0.26\,\%  & $-$0.53\,\%  \\
                      & \textbf{AVG} & \textbf{$-$0.53\,\%} & \textbf{0.17\,\%} & \textbf{0.02\,\%}  & \textbf{$-$0.47\,\%}  \\
    \bottomrule
\end{tabular}
    }
    \label{table_ccso_band_edge_luma}
  \end{center}
\end{table}




The visual quality comparison is presented in Fig.~\ref{visual}. In column~(a), the source frame of the video sequence is shown. Column~(b) highlights the regions where CCSO is applied. Column~(c) shows the specific regions selected from sequences with CCSO disabled, whereas column~(d) displays the same regions from sequences with CCSO enabled. From column~(b), it is observed that the rate of applying CCSO is very high over the whole picture. Comparing columns~(c) and (d), it is evident that the application of CCSO successfully enhances and restores the textures of chroma components. The notably sharper edges in column~(d) illustrated the effective of EO classifier, whereas the detailed textures show the strength BO classifier. In general, it is concluded from the analysis that, the proposed CCSO method recovers more details in the reconstruction image.

\section{Conclusion}
\label{sec:conclusion}

A Cross-Component Sample Offset approach is proposed and implemented on top of a developing AOMedia video codec towards a next-generation video coding standard beyond AV1. The proposed method is featured as a non-linear multiplication-free loop filtering method that explores cross-component statistical redundancy to achieve higher coding efficiency and better subject quality. Simulation results show that, under the common test conditions specified by AOMedia, the proposed method achieved significant objective BD-rate coding gain for several mainstream quality metrics, such as PSNR and VMAF, and observable subjective quality enhancement, with marginal increased complexity and hardware friendly decoder implementation.

\bibliographystyle{IEEEtran}
\bibliography{IEEEabrv, ccso_tip}

\end{document}